\documentclass{aa}
\usepackage{psfig,astron}

\newcommand{\Msun}{\mbox{\rm M$_{\odot}$}}

\begin{document}
   \thesaurus{04     
              (10.08.1;  
               10.06.1;  
               10.19.2;  
               12.04.1;  
               10.07.1)} 
\title{Early galaxy evolution from deep wide field star counts}
\subtitle{I. The spheroid density law and mass function}
\author{A.C. Robin 
\inst{1} 
\and 
C. Reyl\'e 
\inst{1} 
\and 
M. Cr\'ez\'e 
\inst{2}}

   \offprints{A.C. Robin}
	\mail{Annie.Robin@obs-besancon.fr}

   \institute{CNRS ESA6091, Observatoire de Besan{\c c}on, BP1615, 
    F-25010 Besan{\c c}on Cedex, France\\
    email:annie.robin@obs-besancon.fr, celine.reyle@obs-besancon.fr
	\and
   Universit\'e de Bretagne-Sud, F-56000 Vannes, France\\
	email:michel.creze@univ-ubs.fr}

   \date{Received ; accepted }
\titlerunning{The spheroid density law and initial mass function}
\authorrunning{Robin et al.}
\maketitle

\begin{abstract}
As part of a global analysis of deep star counts to 
constrain scenarii of galaxy formation and evolution, we investigate
possible links between the galactic spheroid and the dark matter halo.
A wide set of deep star counts at high and intermediate 
galactic latitudes is used to determine the large scale density law 
of the spheroid. 
Assuming a power density law,  
the exponent, flattening, local density and IMF slope of this population are estimated. 
The estimation is checked for robustness against contamination of star counts by the thick 
disc population. Contamination effects are derived from a model of population
synthesis under a broad variety of thick disc parameters.
The parameter fit is based on a maximum likelihood criterion. 
The best fit spheroid density law has a flattening of 0.76,
a power index of 2.44. There is a significant degeneracy between
these two parameters. The data are also compatible with
a slightly less flattened spheroid
(c/a = 0.85), in combination with a larger power index (2.75). 
A flatter spheroid (c/a = 0.6) with a power
index of 2 is not excluded either. 
We also constrain the spheroid IMF slope $\alpha$ to be 1.9 $\pm$ 0.2,
leading
to a local density of 1.64  10$^{-4}$ stars pc$^{-3}$ and
a mass density of 4.15 10$^{-5}$ \Msun pc$^{-3}$.
With this slope the expected mass density of brown dwarfs in the halo
makes
a negligible part of the dark matter halo, as already estimated from
microlensing surveys.
 
So, as star count data progresses in depth and extent,
the picture of the spheroid star population that comes out points to a shape
quite compatible with what we know about the distribution of  
baryonic dark matter if it is made of stellar remnants, suggesting a 
common dynamical origin.

\end{abstract}

\section{Introduction}

This paper is part of a global analysis of star counts developed to constrain
consistently scenarii of galaxy formation and evolution.The central tool of this approach
is the "Besan\c{c}on" model of population synthesis. This model is gradually tuned
to fit an increasing number of observational constraints while keeping compatibility
with previous fits and theoretical prescriptions. 
In the present paper we address the problem of the halo (dark or visible)
by trying to compare the properties of the spheroid population (the visible
halo) with the dark matter halo, as traced by microlensing at high galactic
latitudes and by the rotation curve. If the dark matter is made at least
partly of stellar remnants, as shown by recent statistics of
microlensing at high galactic latitudes
 \cite{Aubourg1993Natur.365..623A,Alcock1997ApJ...486..697A,1998ApJ...499L...9A},
 the density trend of this matter should be close to a power law 
with index of 2 
(as expected from a flat rotation curve). It is natural to 
think of a similar shape for the stellar spheroid. 

Constraints on the overall shape of the dark halo are poor.
Cosmological simulations of halo formation generally predict that halos 
are flattened
by about c/a $\sim$ 0.7 \cite{1996IAUS..169...23R}. 
But the axis ratio depends on how 
much the halo matter is dissipative, the more 
dissipative, the flatter the halo. Direct determinations of the dark matter distribution in
polar ring galaxies show flattened halos with c/a $\sim$
0.5 \cite{sackett94,1996IAUS..169...23R}.

Concerning the spheroid population,
most previous
analyses suggest rather steep density slopes with power indices
between 3.0 and 3.5. However, these analyses are based on rather small 
samples of well identified tracers. The estimated
flattening also cover a wide range between 0.6 to 1.0:

The distribution of galactic globular clusters appears to be well fitted 
by a power law density with index $n \sim 3.5$ and flattening of 1. 
\cite{harris76,zinn85}. Hawkins RR Lyrae  
observations \cite*{hawkins84}  showed $n = 3.1 \pm 0.2$ with a 
flattening of 0.9. Saha \cite*{saha85}, 
using a spherically symmetric model, found $n \sim 3$ out to 25 kpc but then 
the RR Lyrae density falls off more rapidly beyond 25 kpc. Another study of 
RR Lyrae by Wetterer \cite*{wetterer96} showed that a spherically symmetric model 
yields $n \sim 
3$ whereas an ellipsoidal distribution yields $n \sim 3.5$. Sluis \cite*{sluis98} 
counted 
blue horizontal branch (BHB) stars and RR Lyrae and found $c/a \sim 0.5$ and 
$n = 3.2 \pm 0.3$. Still from BHB star counts, Sommer-Larsen \cite*{sommer87} derived 
$c/a \sim 
0.8$ and $n \sim 3$ up to 40 kpc, Preston \cite*{preston91} found that $c/a$ 
increases from 0.5
to 1 up to 20 kpc with $n = 3.5$. Soubiran \cite*{soubiran93} showed that 
$n = 3.5 \pm 0.5$ is
compatible with the kinematical behavior of a star sample near the 
north galactic pole. K dwarf
counts with HST yield $c/a = 0.8 \pm 0.1$ and $n = 3.06 \pm 0.22$ 
\cite{gould98}.
All of these studies were based on a few hundred objects at most.

In order to find new constraints on the spheroid density law, we  
undertook a photometric and astrometric sample survey in various 
galactic directions. We complemented these
data with existing deep photometric star counts in several high and 
intermediate
latitude fields. Most such counts contain large numbers of halo dwarfs, but
they cannot be 
distinguished from thick disc dwarfs by their colours but at faint magnitudes. 
Since no large optical surveys were
available at magnitude fainter than 20, we used heterogeneous data coming
from various studies (often of extragalactic aim) in various 
photometric systems. 

The population synthesis model used here permits to perform a global analysis
 of these heterogeneous data, since observational data can be simulated in each 
field with the true observational conditions
(photometric system, errors and selection effects). The synthetic approach allows also to
estimate the biases and expected contaminations by other populations.

In section~2 we describe the model of population synthesis and external
constraints on the spheroid population.
In section~3 we describe the data sets and the comparison method.
In section~4 we discuss the results and their implications for the dark matter halo.

\section{The model of population synthesis}

We have used a revised version of the Besan\c{c}on model of 
population synthesis.
Previous versions were described in Bienaym\'e et al. 
\cite*{Bienayme1987A&A...180...94B,Bienayme1987A&A...186..359B} and Haywood 
et al. \cite*{Haywood1997A&A...320..440H}. 

The model is based on a semi-empirical approach, where physical constraints
and current knowledge of the formation and evolution scenario of the
Galaxy are used as a first approximation for the population synthesis. 
The model involves 4 populations (disc, thick disc, halo and bulge) each deserving a specific treatment. 
The bulge population which is irrelevant for this spheroid analysis will be described elsewhere.

\subsection{The disc population}

A standard evolution model is used to produce the disc population,
based on a set of usual parameters : an initial mass function (IMF), a star
formation rate (SFR), a set of evolutionary tracks (see Haywood et al., 1997
and references therein). 
The disc population is assumed to evolve during 10 Gyr. 
A set of IMF slopes and SFR's are tentatively assumed and tested against
star counts. The tuning of disc parameters  against relevant 
observational data was described in Haywood et al.
\cite*{Haywood1997A&A...320..428H,Haywood1997A&A...320..440H}.

The model fixes the distribution of stars in the space 
of intrinsic
parameters : effective temperature, gravity, absolute
magnitude, mass and age. These parameters are converted into
colours in various systems through stellar atmosphere models 
corrected to fit empirical data 
\cite{Lejeune1997A&AS..125..229L,Lejeune1998A&AS..130...65L}. While some errors
still remain in the resulting colours for some spectral types, the overall
agreement is good in the major part of the HR diagram.

Since the Haywood et al. model was based on evolutionary tracks at solar
metallicities, inverse blanketing corrections are introduced to give to the
disc a metallicity distribution in agreement with Twarog \cite*{Twarog80} 
age/metallicity distribution (mean and dispersion about the mean).

 The model returns the present-day distribution of stars as a function of 
intrinsic parameters in a unit volume column 
centered at the sun position. Since the 
evolution model does not account for orbital evolution, stars are
redistributed in the reference volume over the z axis. 
The key for redistributing stars along the z-axis is age : an empirical
relation associates z velocity dispersions to ages. Then the Boltzmann 
equation is used to convert z velocity distributions into z density. 
The model is dynamically self-consistent in the sense that the potential 
used in the Boltzmann equation 
is the one generated by the total mass distribution of stellar populations.
The self consistency is established iteratively.   
We slice the disc populations into seven isothermal populations
of different ages, from 0 to 10 Gyr. Each sub-population (except the 
youngest one, which cannot be considered as relaxed) has its velocity
dispersion imposed by the age/velocity dispersion relation. We then deduce 
the scale height of each sub-population using
the Boltzmann equation. The overall scheme is described in Bienaym\'e \cite*{Bienayme1987A&A...180...94B}.

Resulting density laws are used to correct the evolution model distribution
in and off the plane, then to compute the stellar densities all over the
Galaxy.

\subsection{The thick disc population}

A detailed analysis of the thick disc population from
photometric and astrometric star counts has been given elsewhere 
\cite{Ojha1994A&A...284..810O,Ojha1994A&A...290..771O,Ojha1996A&A...311..456O,Robin1996A&A...305..125R,1999A&A...351..945O}. 
The kinematics, metallicity,
and density law were measured allowing us to constrain
the origin for this population. In this series of papers, evidence was given that the majority of
thick disc stars should originate from a merging event at the beginning of the
life of the thin disc, after the first collapse. One or several satellite
galaxies have heated the thin disc, then the gas re-collapsed and reformed
a new thin disc \cite{Robin1996A&A...305..125R}.

In the population synthesis process,
the thick disc population is modeled as originating from a single epoch of
star formation. We use Bergbush \& Vandenberg \cite*{Bergbush92} oxygen enhanced evolutionary 
tracks. No strong constraint exists on the thick disc age until now. We 
 assume an age
of 11 Gyr, which is slightly older than the disc and younger than the halo.
The initial mass function is modeled by a simple power law with a slope 
about  $\alpha=1-2$,  
referring to the notation $\phi(m) \propto m^{-\alpha}$. 

The thick disc metallicity can be chosen between \mbox{-0.4} and \mbox{-1.5} dex in the 
simulations. The standard value of \mbox{-0.7} dex is usually adopted, following
in situ spectroscopic determination from Gilmore et al.
\cite*{1995AJ....109.1095G} and photometric star count 
determinations \cite{Robin1996A&A...305..125R,1999A&A...348...98B}. 
The low metallicity tail of the thick disc seems to 
represent
a weak contribution to general star counts \cite{1993AJ....105..539M}.
It was neglected here.
An internal metallicity dispersion among the thick disc 
population is allowed.
The standard value for this dispersion is 0.25 dex. No evidence has been 
found for
a significant metallicity gradient in the thick disc population 
(Robin et al, 1996).
 
The thick disc density law is assumed to be a truncated exponential : at 
large distances
the law is exponential. At short distances it is a parabola. This formula
ensures the continuity and derivability of the density law 
(contrarily to a true
exponential) and eases the computation of the potential. The scale height 
of the exponential can vary between 600 and 2600 pc.
The standard value, 760 pc, has been obtained from star count fitting in 
various directions
(Robin et al, 1996). Nevertheless, it can be shown that star counts 
when restricted to
a small number of galactic directions and a small magnitude range 
do not give a strong constraint on the scale height, but rather on the 
parameter : (local density)$\times$(scale height)$^2$. 
At present there is no accurate determination
of the thick disc density in the solar neighbourhood, independently from
the scale height.
But reasonable values range between 700 to 1200 pc for the scale height
and 1 to 4\% for the local density relative to the thin disc.

\subsection{The spheroid}

We assume a homogeneous population of spheroid stars with a
short period of star formation. We thus use the Bergbush and Vandenberg
\cite*{Bergbush92}
oxygen enhanced models, assuming
an age of roughly 14 Gyr (until more constraints on the age are available), 
a mean
metallicity of -1.7 dex and a dispersion of 0.25 about this value. No 
galactocentric
gradient is assumed. The IMF has to be constrained either from globular
clusters (if they are representative of the spheroid population) or 
from deep star counts. This point is discussed in section~3.

The density of spheroid stars is modeled by a power law :

\begin{center}
\[ \rho(R,z) = \rho_0 \times (R^2+\frac{z^2}{\epsilon^2})^{n/2}\]
\end{center}

where $\rho_0$ is the local density,
$n$ is the power law index and $\epsilon$ is the flattening.

The local density can be constrained by local measurements 
of high velocity
stars, or by remote counts of giants (spectroscopically selected) or
dwarfs (photometrically selected). The local density cannot be 
determined independently
from the other density parameters with our limited number of data sets. 
Thus we have used independent constraints from the literature 
on the local spheroid density. 

\subsection{The local spheroid density}

\begin{figure*}
\vspace{0cm}
\hbox{\hspace{0cm}\psfig{figure=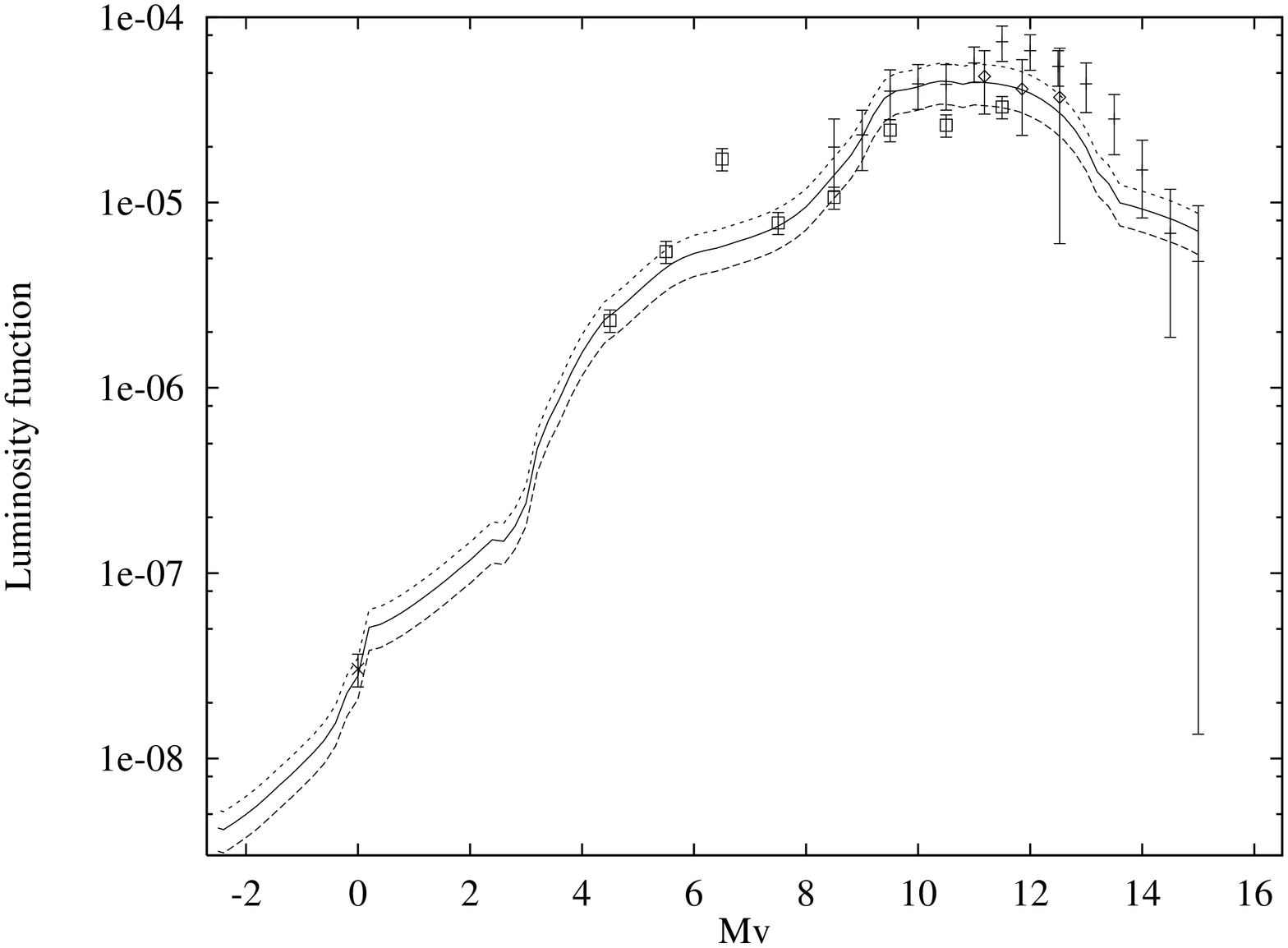,width=9cm,angle=-90}\hspace{0cm}
\psfig{figure=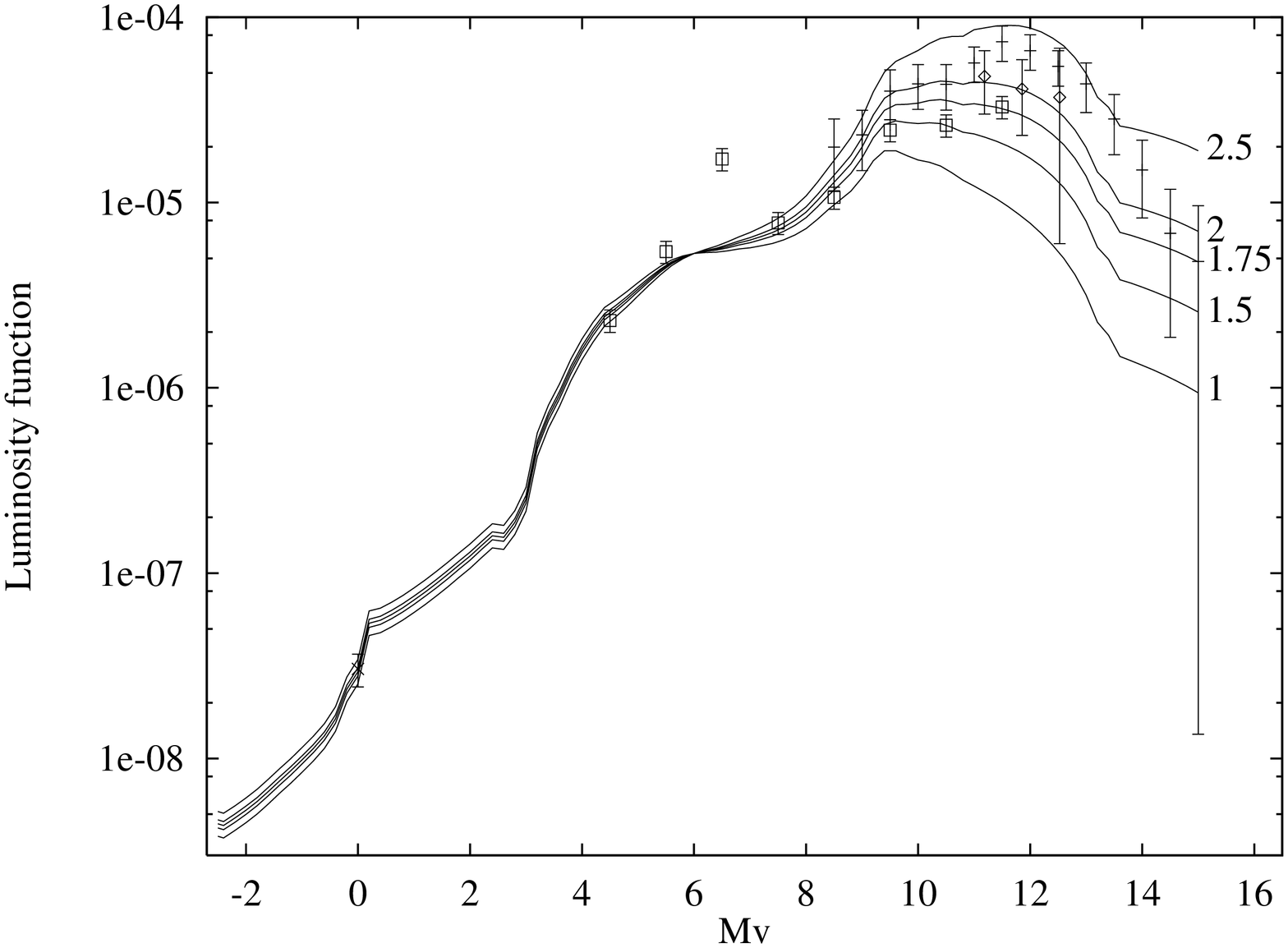,width=9cm,angle=-90}}
\vspace{0cm}
\caption[]{Luminosity function of the stellar halo. \cite{gizis99}: diamonds;
\cite{bahcall86b}: squares; \cite{dahn95}: plus, \cite{morrison93}: cross.
a: solid line: modeled luminosity function with an IMF slope ($\alpha$) of 2.0;the dotted lines give the same luminosity function
renormalized by a factor 0.75 or 1.25. b: luminosity functions as a function
of IMF slope (as indicated in the graph).}
\label{fig-flum}
\end{figure*}

The local stellar spheroid density, $\rho_0$, is bounded by observational 
data 
on halo dwarfs and giants. Figure~\ref{fig-flum} shows the luminosity 
function 
obtained by different authors. We only selected recent results 
obtained in good conditions from sufficiently large samples. 
Bahcall et al \cite*{bahcall86b} and Gizis \& Reid \cite*{gizis99} derived 
their values from high proper motion dwarf samples. Dahn et al \cite*{dahn95} 
determined 
accurate parallaxes for local late-type subdwarfs and deduced the local
luminosity function of halo stars in the absolute visual magnitude range 
9 to 14. These three results are biased by the kinematic selection. They
took the bias correction into account but this correction is model
dependent and introduces an unknown uncertainty into the result. We expect that
the differences between the three measurements rely upon this correction.
On the giant side, Morrison \cite*{morrison93} used a 
non kinematically-biased sample of halo giants, selected from their 
metallicity to estimate the spheroid local density. In figure~\ref{fig-flum}a 
we show the luminosity function from Bergbush and Vandenberg (1992) for
a population of 14 Gyr with a metallicity of -1.75 and an 
IMF slope $\alpha$ of 2. If we 
let the local halo density vary from a factor of 0.75 to 1.25 relative 
to this reference model (dotted lines in figure~\ref{fig-flum}a), 
we get a good agreement 
with the specified observations given their uncertainties. In the next section
we allow the local density to vary within these limits.

\section{Data sets and fitting methods}

Obtaining good constraints on the spheroid density law requires 
a good photometric accuracy. This generally depends on using CCD detectors
on large telescopes, on fields as wide as possible to cover large samples,
and a large range of galactocentric distances.
This can be obtained with a number of data sets at various galactic
latitudes and longitudes. We have collected such data sets from the 
literature. Most have been made for extragalactic purposes.

\subsection{Available data}

The main data characteristics are summarized in table~1.
The photometric systems are  close to the Johnson-Cousins system. 
Spheroid star selection was based on their magnitude and colour 
(either B-V, V-R, or V-I, depending on the available observations), 
in order to avoid presence of contamination by other populations.
 Aiming at model independent results, the model was used essentially to 
select 
colour and magnitude ranges and fields where the contamination by thick 
disc stars
remains negligible under any reasonable thick disc hypothesis. For 
this reason
all data brighter than magnitude 20 at intermediate and low latitudes were
excluded.
A small number of disc white dwarfs is also present
in the selection but the proportion is at most a few percent and has no 
consequence on the result.

Our survey program include at the moment two fields, one towards the north
galactic pole, another at intermediate latitude (l=150,b=60). The 
NGP field
is the deepest up to now : it is complete and free from galaxy contamination
up to magnitude 24. A full description of these data sets will be given 
in a forthcoming paper.

The other selected data sets are the six fields of the DMS survey 
\cite{1996ApJS..104..185H,1998ApJS..119..189O}
observed in V and R bands at medium latitude, 4 fields
from the Canada-France Redshift survey (CFRS, \cite{1995ApJ...455...50L,1995ApJ...455...60L,1995ApJ...455...88H}) dedicated to galaxy counts, two fields 
from the 
Koo and Kron investigation for quasars \cite{Koo82,Koo86}, another field from 
Reid and Majewski near the north galactic pole 
\cite{Reid1993ApJ...409..635R}.

\begin{figure}
\vspace{0cm}
\hspace{0cm}\psfig{figure=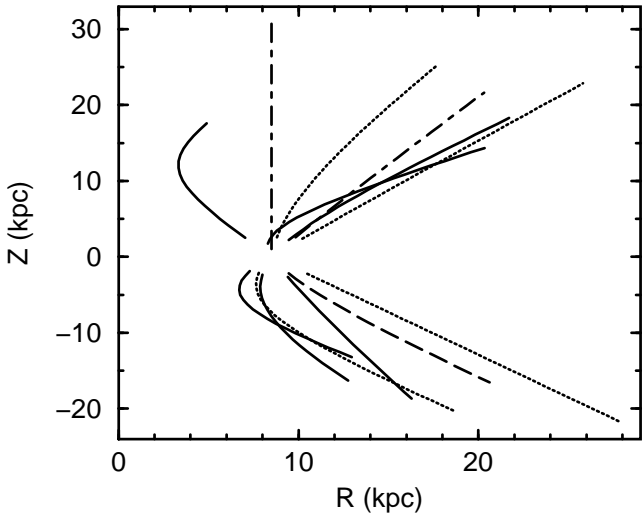,width=8.8cm,angle=-90}
\vspace{0cm}
\caption[]{Line of sight projected on (R,z) plane. The segments limits 
indicate the distance
distribution for 90 \% of the halo stars. Solid lines indicate the 
Deep Multicolor
Survey fields, dotted lines the Canada-France Redshift Survey 
fields, long dashed lines the field from Koo \& Kron \cite*{Koo82} and the dot-dashed lines 
the fields from our program, as well as other North Galactic pole fields
from Koo et al. (1986) and Reid et al. \cite*{Reid93}. }
\label{fig-Rz}
\end{figure}

The absolute visual magnitude of halo stars in the selected samples ranges 
between 3 and 8, except our north galactic pole field which reaches 
M$_V \sim 11$. 
All these fields taken together cover a large part of the (R,z) plane, 
as can be seen in figure~\ref{fig-Rz} where the 
distributions in R and z of 90\% of halo stars in each field of view are drawn.

\begin{table*}
\begin{center}
\caption{Deep photometric surveys used in our analysis. The 
magnitude and colour range
used to select the halo stars are given.\label{tab1}}
\begin{tabular}{llllll}
\hline
Reference 	& Field                   & Area     & Bands  & Magnitude 
			& Colour\\
          	& coordinates             & (deg$^2$)&        &  range    
			& range\\
\hline\hline
Our program   	& North Galactic Pole     & 0.158    & V,I    & V=20,24     
			& 0,1.6\\
	     	& l=150$^o$, b=+60$^o$    & 0.051    & V,I    & V=20,22     
			& -0.4,1.2\\
DMS   & l=129$^o$, b=-63$^o$    & 0.144    & V,R    & V=20,22     
			& 0,0.6\\
 		& l=248$^o$, b=+47$^o$    & 0.079    & V,R    & V=20,22     
			& 0,0.6\\
   		& l=337$^o$, b=+57$^o$    & 0.156    & V,R    & V=20,22     
			& 0,0.6\\
		& l=77$^o$, b=+35$^o$     & 0.153    & V,R    & V=20,22     
			& 0,0.6\\
		& l=52$^o$, b=-39$^o$     & 0.149    & V,R    & V=20,22     
			& 0,0.6\\
		& l=68$^o$, b=-51$^o$     & 0.149    & V,R    & V=20,22     
			& 0,0.6\\
Koo \& Kron (1982)    & North Galactic Pole     & 0.097    & J,F    & J=20,22     
			& -0.4,1.0\\
		& l=111$^o$, b=-46$^o$    & 0.299    & J,F    & J=20,22     
			& -0.25,1.0\\
Reid \& Majewski (193)   & North Galactic Pole     & 0.300    & V,B    & V=20,22     
			& 0,0.8\\
CFRS      	& l=177.4$^o$, b=-48.3$^o$&0.0105    & V,I    & I=20,22     
			& -0.4,1.4\\
	& l=205$^o$, b=52$^o$     &0.0033    & V,I    & I=20,22     
				& -0.4,1.4\\
	& l=96.3$^o$, b=59.9$^o$  &0.00952   & V,I    & I=20,22     
			& -0.4,1.4\\
	& l=64$^o$, b=-44.4$^o$   &0.0062    & V,I    & I=20,22     
			& -0.4,1.4\\
\hline
\end{tabular}
\end{center}
\end{table*}

\subsection{Analysis method} 

Population synthesis simulations have been computed in every observed field
using photometric errors as close as possible to the true observational
errors, generally with photometric errors growing as a function of the
magnitude and assumed to be Gaussian. Monte Carlo simulations are done 
in a solid angle much larger 
than the data in order to minimize the Poisson noise.

Then we compare the number of stars produced by the model 
with the observations in the selected region of the plane (magnitude, colour)
and we compute the likelihood of the observed data to be a realization of the 
model (following the method described in Bienaym\'e et al, 1987a, appendix C).
The likelihood has been computed for a set of models, varying the power
law index between 2.0 and 3.5, the flattening between 0.3 and 1.0, the
local density between 0.5 and 1.25 times the standard value as defined in
section~2.4, and the spheroid IMF slope $\alpha$ from 1.0 to 2.2.

The confidence limits of estimated parameters are determined by the            
likelihood level which can be reached by pure random change 
of the sample :                       
a series of simulated random samples are produced using the set of model
parameters. 
The rms dispersion of the likelihood about the mean of this series gives 
an estimate of the likelihood fluctuations due to the random noise. It 
is then used to compute the confidence limit. Resulting 
errors are not strictly 
speaking standard errors, they give only an order of magnitude.

\section{Results and discussion}

\subsection{Constraints on the spheroid density law}

\begin{figure*}
\vspace{0cm}
\hspace{0cm}\psfig{figure=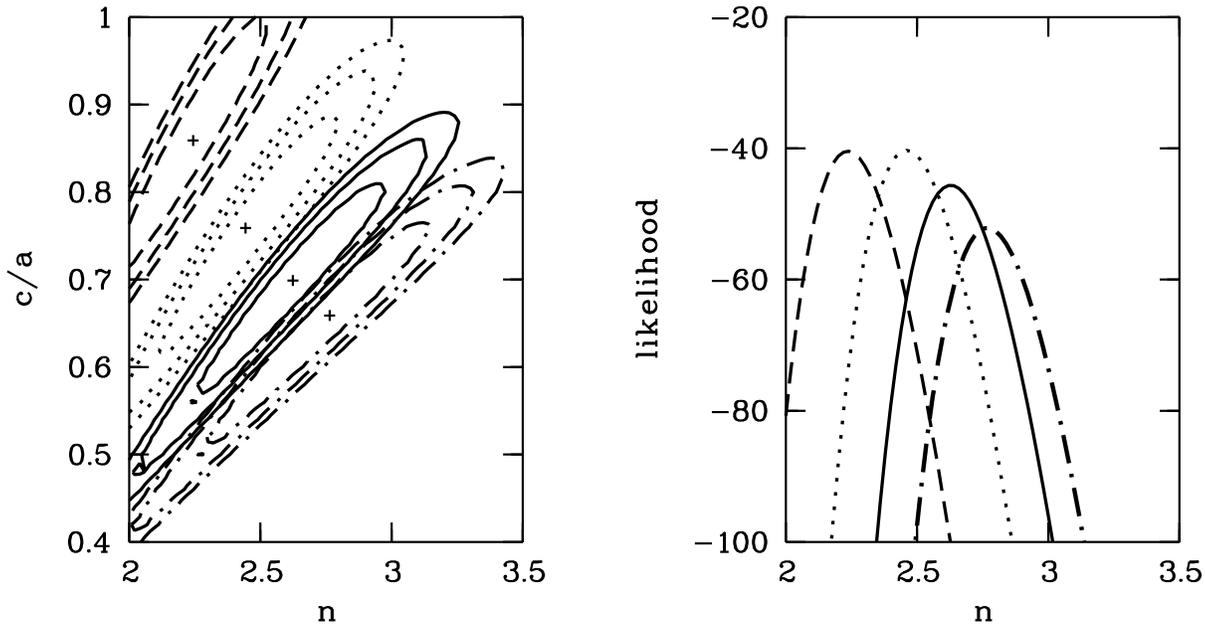,width=17cm,angle=-90}
\vspace{0cm}

\caption{Likelihood as a function of power law index, flattening and local 
density: On the left: iso-likelihood contours at 1, 2 and 3 sigmas 
in the plane (power law index, flattening)
of the spheroid population. On the right, likelihood as a 
function of the power law index, for the 
best flattening. Lines are coded by the local density. Solid line : 
standard local density 
as defined in section~2.4; dotted line: 0.75 $\times\rho_0$; 
dashed: 0.50 $\times\rho_0$, dotted-dashed: 1.25 $\times\rho_0$}
\label{fig-lr}
\end{figure*}

Figure~\ref{fig-lr} gives the value of the likelihood as a function
of the flattening, power law index and local normalization.
On the left, iso-likelihood contours are drawn for four values of the
local normalization (0.5, 0.75, 1.0 and 1.25 $\times \rho_0$). 
On the right, we show the likelihood values as a function of the
power law index for the best fit value of the flattening. 

Comparing the results of different local normalization we 
conclude that the choice of the local normalization 
sensitively displaces the best fit power law index and flattening, but
their likelihood are not similar. The best fit model
is obtained either with a local density of 0.75 $\rho_0$, a power law index
of 2.44 and an axis ratio of 0.76, or a local density of 0.5 $\rho_0$, a 
power law index of 2.24 and an axis ratio of 0.86. The values obtained with
a standard local density $\rho_0$ are slightly worse but stay within
1 sigma confidence level.
They are a power law index of 2.62 and an axis ratio of 0.70. 
The best fit local densities 0.5$\rho_0$ and 0.75$\rho_0$ 
agrees with  the
Bahcall \& Casertano determination of the local luminosity function as seen
in figure~1, but conflict
with Dahn et al., which favors a local density of 1.25 $\rho_0$. 
However, in the 
present study the statistics is dominated by stars with absolute magnitudes in the range 3 to 8,
a range poorly represented in the Dahn et al. sample. Only deeper 
counts could give
constraints on the fainter part of the luminosity function.

\begin{figure*}
\vspace{0cm}
\hspace{0cm}\psfig{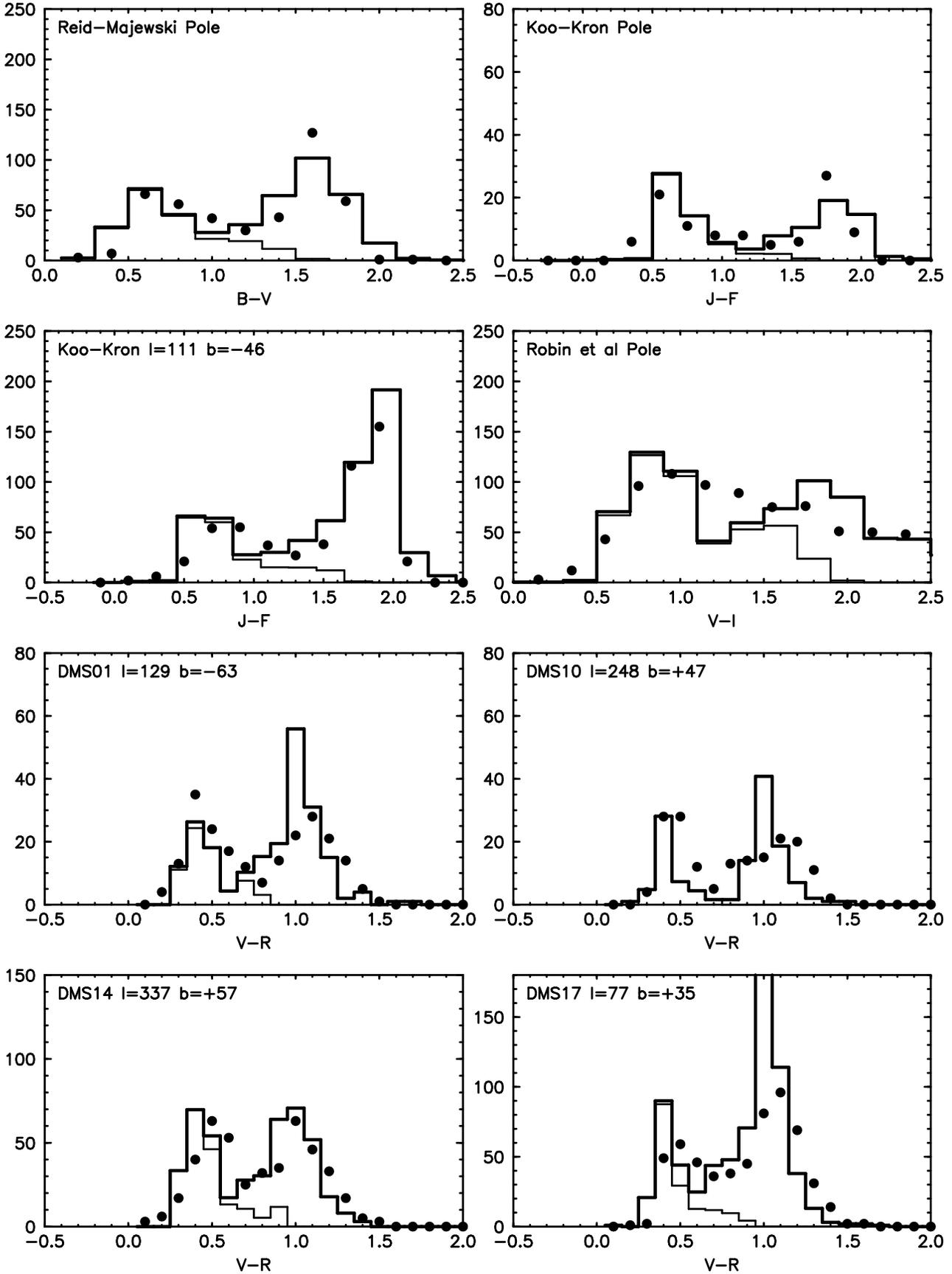}
\vspace{0cm}

\caption{Colour distributions of selected samples. Contributions of 
different magnitude 
interval have been summed up for clarity. Observations in 
different fields are shown by dots. The ordinates give the true
number of observed stars in each colour bin. Heavy solid lines are the predicted number of stars
by the model assuming a power law index of 2.44, a flattening of 0.76 and
a local density of 0.75$\times \rho_0$
for the spheroid. Thin lines show the contribution of the spheroid alone.}
\label{fig-colora}
\end{figure*}

\begin{figure*}
\vspace{0cm}
\hspace{0cm}\psfig{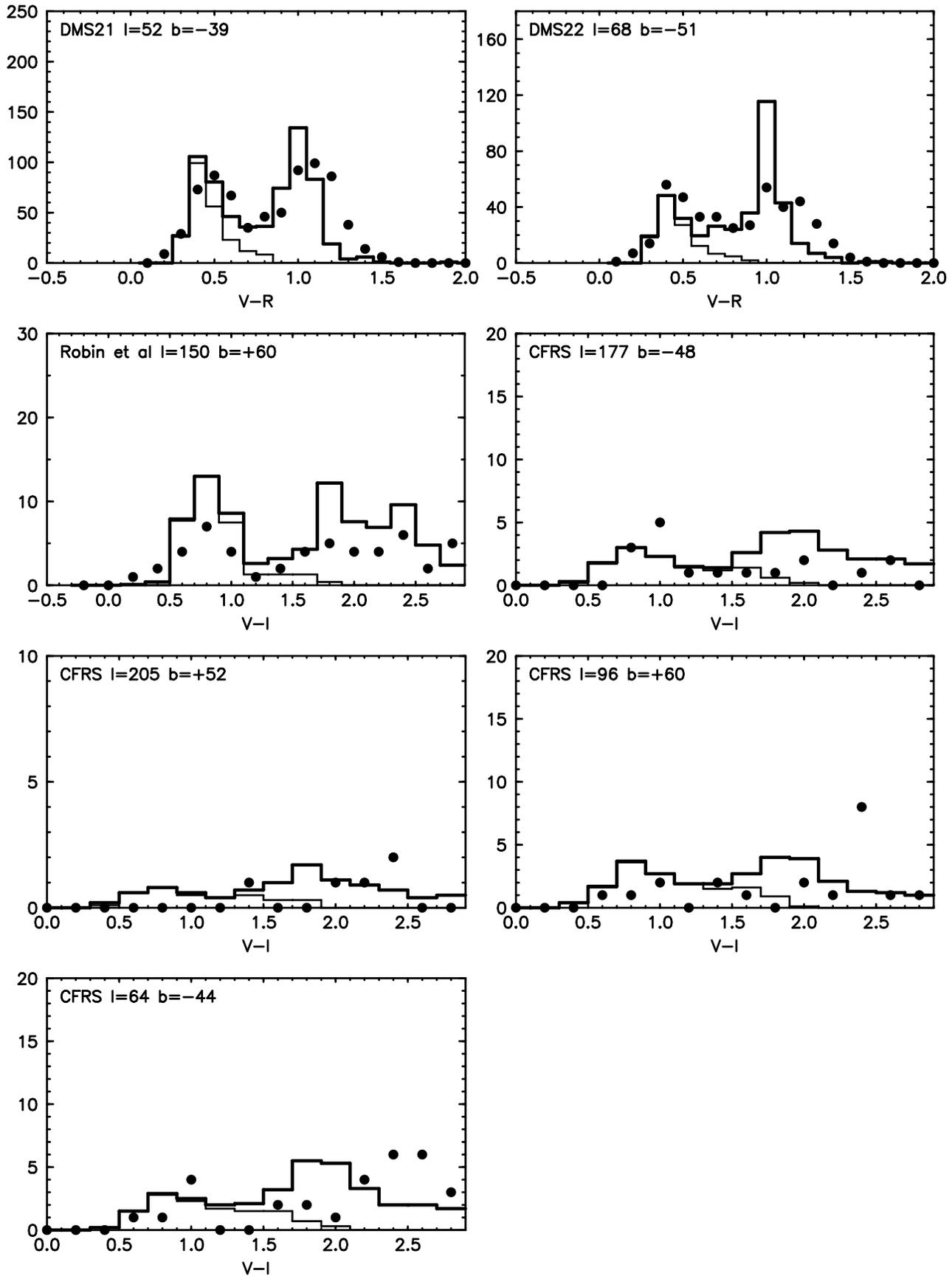}
\vspace{0cm}
\caption{Same as figure~\ref{fig-colora} for other fields}
\label{fig-colorb}
\end{figure*}

It is worth having a look at the colour distributions as predicted by the
best fit model compared with the observational data.
Figure~\ref{fig-colora} and ~\ref{fig-colorb} show the colour 
distributions observed (dots) and 
predicted (heavy solid line) by the
best fit model ($\epsilon$=0.76, n=2.44, 0.75$\rho_0$) in the selected 
magnitude interval in each
tested field. Superimposed we show the distribution of the spheroid population
alone as predicted by the model (light solid line). We see here that some
photometric systems are not closely matched by the model, as seen by
slight shifts between model and data in some cases. But the way we have 
selected
spheroid stars in the blue peak of the distribution cannot introduce a bias
even in case of colour shifts.

\subsection{Sensitivity to the IMF slope}

Whatever the assumed IMF slope in the range 1-2.2, the maximum
likelihood is
obtained for the same density law parameters.
 There is a slight
likelihood variation related to the choice of the IMF, but it is
only due to the deepest magnitude bin towards the pole.

A separate analysis of
star counts deeper than 22 towards the pole can help
determine precisely the IMF slope. In this magnitude range spheroid
stars with
absolute magnitude 10-11 contribute substantially to star counts, while
their
frequency is sensitive to the IMF as can be seen on figure~1b.
The analysis is slightly different from the determination of
the density law. In this range of absolute magnitude the subdwarf sequence
turns redwards making the colour index a good luminosity indicator.
The V-I distribution is used as an additional constraint. A V-I
histogram is built with a bin 0.1 magnitude wide over the
range 0. to 3. The density law is adopted from the above analysis, so
the free parameters are
the halo and thick disc IMF slopes. Since these two populations are quite
well separated in the (V,V-I) plane, the two IMF slope estimates are
de-correlated. Table~2 gives the resulting slope estimates with their
likelihood in the magnitude range 22-24 for the different spheroid
density laws determined previously.

\begin{table}
\begin{center}
\caption{IMF slopes of the spheroid and thick disc populations determined
by the maximum likelihood technique for four models of spheroid density
laws}
\begin{tabular}{cccccc}
\hline
 $\rho_0$ & $n$ & $\epsilon$ & $\alpha$ Spheroid  & $\alpha$ Thick disc
& Likelihood \\
\hline
0.50 & 2.24 & 0.86 & 1.9 & 1.7 & -79.1\\
0.75 & 2.44 & 0.76 & 1.9 & 1.7 & -58.8\\
1.00 & 2.62 & 0.70 & 1.9 & 1.6 & -55.9\\
1.25 & 2.76 & 0.66 & 1.8 & 1.6 & -60.3 \\
\hline
\end{tabular}
\end{center}

\end{table}

Spheroid models with a local density 0.75 $\rho_0$ and 1.0 $\rho_0$ give
the maximum likelihood, well in agreement with the previous result.
However the model with 0.50 $\rho_0$ is noticeably worse. Eventually, the
resulting IMF slopes do not depend significantly on the assumed 
density laws and the
likelihood is well peaked around the maximum indicating a robust
determination.
We conclude that the IMF of the halo, in the mass range [0.1, 0.8] is :

\[ \phi(m) \propto m^{-1.9} \]

\noindent while the IMF slope of the thick disc seems to be slightly smaller
and similar to the disc's \cite{Haywood94}.
These values do not account for binarity. Thus the true IMF should slightly
steepen. We leave the value uncorrected until more data are available on
the binary fraction among low mass spheroid stars.

This result is the first direct measurement of  the mass function of
field star spheroid with a good statistics, thanks to the wide field of
the CCD mosaic. Several previous determinations  used
kinematically selected samples (see section~2.4 for references) or
deep fields. But the latter were limited to narrow fields : the first attempt
by Richer \& Fahlman \cite*{Richer92} lead
to a very steep IMF slope of $\alpha=4.5\pm1.2$ which had given
hope for a dark matter halo of brown dwarfs. Later results have given
shallower slopes but the uncertainties were not significantly decreased.

Gould et al. \cite*{gould98} analyzed a sample of 166 stars in 53
HST WFPC fields, making difficult the de-correlation between
structural parameters
of the spheroid and its mass function. They found a luminosity function
down by a factor two from the present one and deduced an IMF slope of
$\alpha=0.75$ (in our notation). Their result relies upon the assumption
that
the spheroid has a mean metallicity of -1.0, which looks too high
considering most direct measurement of its abundances. This high metallicity
induces an overestimate of the luminosity at a given colour, hence of the
distance, as well as an overestimate of the mass relatively to a smaller
assumed metallicity.

\subsection{Variations of density law with galactocentric position}

If we independently check the results obtained in inner
fields and in outer fields, we are able to search for solutions with
varying power law index and flattening over the galactic radius.
Contrarily to Preston et al. (1991) we find no evidence for 
varying power law index or flattening. 
However, a round spheroid is ruled out by the 
inner field data as well
as by fields at low latitudes. Thus our results are
compatible with a true power law and a constant flattening all along the 
tested galactic radius.
 
When comparing data sets from different sources in close galactic
fields, discrepancies appear
which are larger than expected on the basis of pure random noise. This may be
due either to data incompleteness, or to 
systematic errors in the photometry (including
mismatch of the standard photometric system), or to true inhomogeneities in
the spheroid distribution. Currently available data are not sufficient 
to discriminate between these different causes. Homogeneous wide field
surveys will be necessary to clarify these aspects. The scope of the current
investigation is for this reason limited to large scale average 
characteristics.

\subsection{Contamination by other populations}   

The blue peak at these magnitudes may be contaminated by disc
white dwarfs or by thick disc main sequence stars. The former are very few
compared to the density of the halo. The contamination by disc white
dwarfs, as determined by
the model, is at most 5\% in the magnitude range 22-24. 

\begin{figure*}
\vspace{0cm}
\hspace{0cm}\psfig{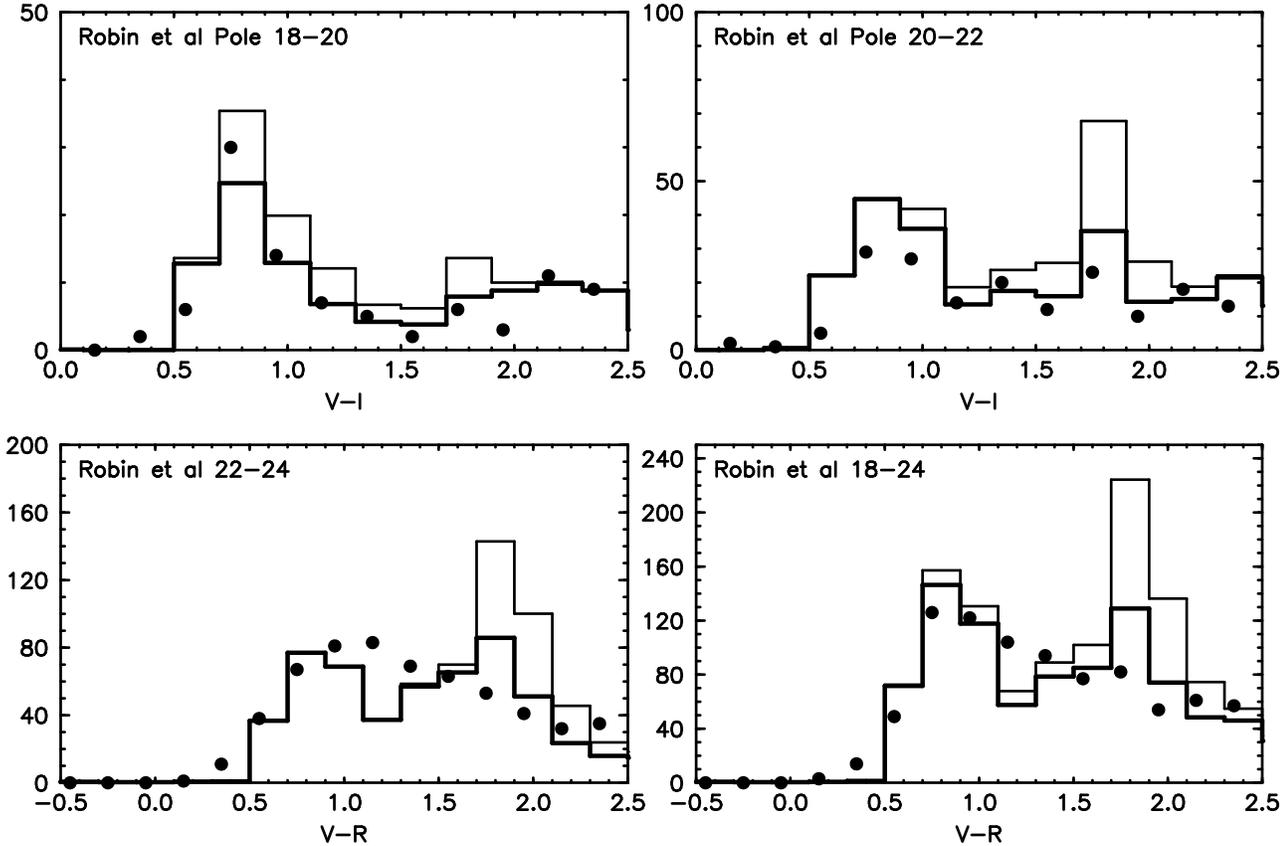}
\vspace{0cm}
\caption{Colour distributions towards the galactic pole in three magnitude
intervals. Dots: observational counts; heavy solid lines: model
distribution assuming a standard thick disc, light lines: model assuming
a doubled thick disc density. The contamination of the blue peak by the 
thick disc can reach
30\% at magnitudes 18-20 but becomes negligible at fainter magnitudes.}
\label{fig-contamin-mag}
\end{figure*}

The contamination
from the thick disc has been estimated using our best fit thick disc model
as adjusted on medium deep star counts. The contamination can reach 
about 30\% in the magnitude range 18-20 but becomes negligible at magnitude 
larger than 20 as seen in figure~\ref{fig-contamin-mag}. Hence, we do not 
take into account magnitudes lower than 20 in our study.

\begin{figure*}
\vspace{0cm}
\hspace{0cm}\psfig{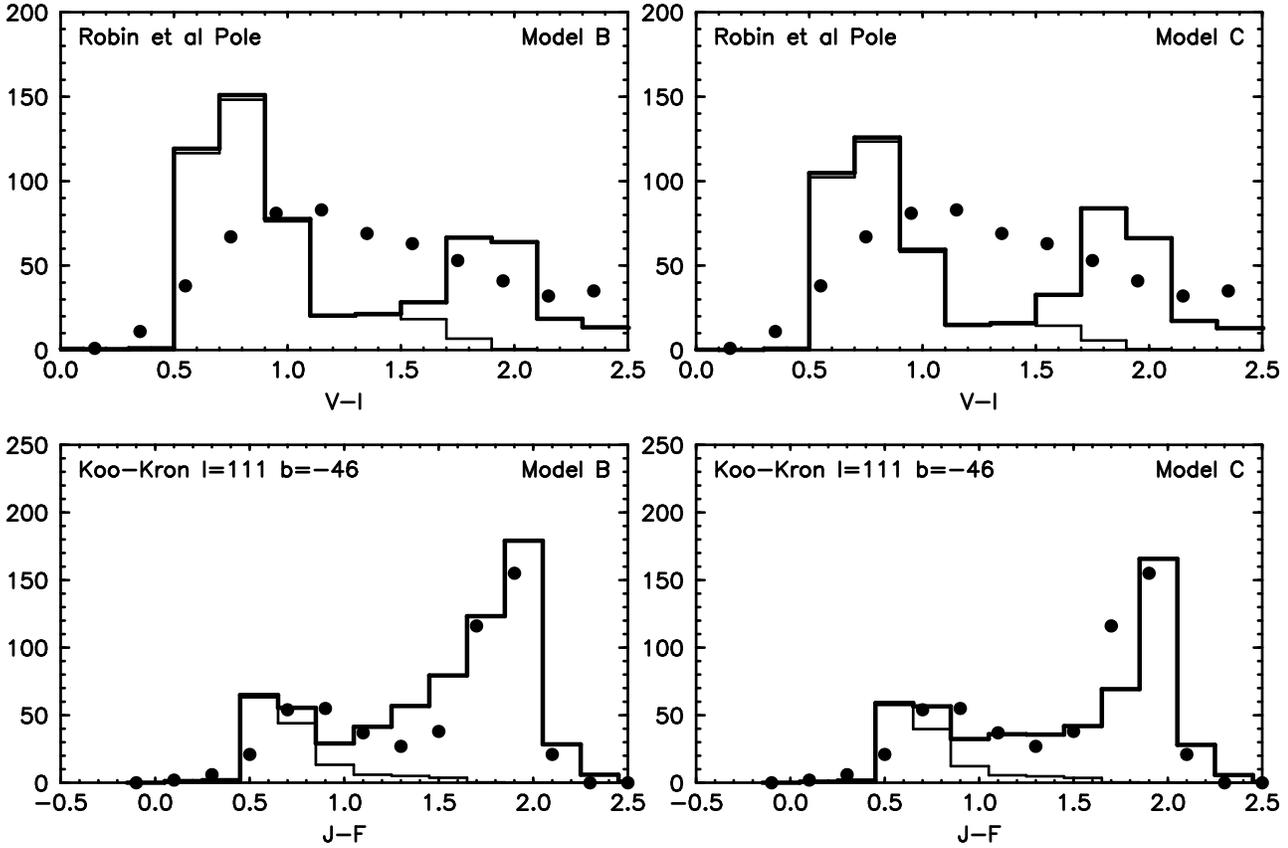}
\vspace{0cm}

\caption{Observed colour distributions towards the north galactic pole (magnitude range 22-24) and
SA68 Koo \& Kron data (magnitude range 20-22) compared with model 
predictions with two thick disc models. On the left side, model B (local 
density of 3.9 \% of the thin disc,
a scale height of 1150 pc and an IMF slope of $\alpha=1$), On the right side, 
model C (local density of 0.5\%, scale height 2 kpc, IMF slope 1.75). 
Thin lines are the contribution of the adjusted spheroid (see text).}
\label{fig-contaminBC}
\end{figure*}

Had the thick disc contamination been underestimated, then the
contribution assigned to the halo in the blue peak would be too large, 
resulting in a possible distortion of the density law.
In order to evaluate how this would affect our 
conclusions, we have investigated different thick disc models which 
could fake the
halo contribution to the blue peak. Attempts were limited
to realistic thick discs roughly fitting the red peak.
 We have selected two extreme thick disc
parameters for which the contamination to the blue peak becomes significant.
A thick disc with a local density of 3.9\%, a scale height of 
1150 pc and an IMF slope of 1. (referred to as model B).  A thick disc 
with a scale height
of 2 kpc, a local density of 0.5\% and an IMF slope of 1.75. (model C).
With such thick disc models, the process of adjusting the spheroid density
law parameters end up to tiny local density of about 25\% of the standard
value and very small power law index of the order of 1.5. Surprisingly,
the fit is good on star counts up to magnitude 22, showing that a large range
of parameters can reproduce a wide set of star counts. However at magnitude
22-24, model B and C are unable to reproduce the counts, as exemplified 
in figure~\ref{fig-contaminBC}, where at the top star counts at the pole in the magnitude 
range 22-24 are overestimated in the blue peak, and in the range 18-22
(bottom) the fit is still acceptable.

 So, the degeneracy between thick disc models and halo parameters holds
only if star counts are not deep enough. Keeping
reasonable values for the thick disc parameters leads to a small contamination
with no risk of underestimation of the halo density. If the thick disc
contribution is higher than expected from standard models then we would 
overestimate the local halo density and power law index, strengthening our
conclusion towards a flat spheroid with a small power law index.

\section{Conclusions}

We have shown that, with the data available up to now, the spheroid
star distribution follows a power law with an index smaller than 
previously thought. It is
moderately flattened, as already
found by several investigations. The best fit power law index is found 
2.44 for a flattening of 0.76. We cannot exclude
a spheroid population having a power law as low as 2 and an axis ratio
of 0.5 at the 2 sigma level.
Assuming a high exponent of 3.5, as suggested by some globular cluster 
and RR Lyrae data, model predictions deviate significantly from observations 
in the external parts of the galaxy whatever reasonable flattening is 
adopted. 

The IMF slope of the spheroid is found to be $\alpha=1.9\pm0.2$, value
which gives a local density of 1.64  10$^{-4}$ stars pc$^{-3}$ and
a mass density of 4.15 10$^{-5}$ \Msun pc$^{-3}$ for the
stellar halo, yet this value ignores possible old white dwarfs. With
this
slope the expected mass density of brown dwarfs in the halo makes a
negligible
part of the dark matter halo, as already estimated from microlensing
surveys.

Recent searches for ancient halo white dwarfs have given a hope to
identify the microlensing events with such objects 
\cite{1999ApJ...524L..95I,Ibata2000,2000Natur.403...57H}. 
Ibata et al. \cite*{Ibata2000} conclude that old white dwarfs may constitute
a significant fraction of about 10\% of the dark matter halo. In the mean time,
microlensing experiments have narrowed the range of the estimated 
halo baryonic 
fraction to 20 to 50\% \cite{alcock2000}. These two results are well 
in agreement according to the uncertainties. 

So, as star count data progresses in depth and extent,
the picture of the spheroid star population that comes out points to 
characteristics 
quite compatible with what we know about the distribution of  
baryonic dark matter if it is made of stellar remnants, suggesting a 
common dynamical origin.
The visible spheroid and its heavy counterpart of dark remnants can
make a significant but not dominant part of the so-called dark matter halo.


\end{document}